\documentclass{sig-alternate-10pt}

\usepackage{indentfirst}
\usepackage{graphicx}
\usepackage{url}
\usepackage{subfigure}
\usepackage{threeparttable}
\usepackage{booktabs}
\usepackage{colortbl}
\usepackage{xcolor}
\usepackage{caption}
\usepackage[linesnumbered,ruled,vlined]{algorithm2e}
\usepackage{algpseudocode}
\definecolor{mygray}{gray}{.9}

\newcommand{\reffig}[1]{Fig. \ref{#1}}
\newcommand{\refsec}[1]{Sec. \ref{#1}}
\newcommand{\reftab}[1]{Table \ref{#1}}

\newcommand{\refalg}[1]{Algorithem \ref{#1}}
\begin{document}
%
\title{We Can Track You If You Take the Metro: Tracking Metro Riders Using Accelerometers on Smartphones}

%
%
%


\numberofauthors{3} 
%
\author{
\alignauthor
Jingyu Hua\titlenote{email: huajingyu2012@gmail.com}\\
       \affaddr{Nanjing University}\\
       \affaddr{Computer Science and Technology}\\
       \affaddr{Nanjing, China}
\alignauthor
Zhenyu Shen\titlenote{email: shenzhenyuyuyu@gmail.com}\\
       \affaddr{Nanjing University}\\
       \affaddr{Computer Science and Technology}\\
       \affaddr{Nanjing, China}
\alignauthor
Sheng Zhong\titlenote{email: sheng.zhong@gmail.com}\\
       \affaddr{Nanjing University}\\
       \affaddr{Computer Science and Technology}\\
       \affaddr{Nanjing, China}
}

\maketitle
\begin{abstract}
Motion sensors (e.g., accelerometers) on smartphones have been demonstrated to be a powerful side channel for attackers to spy on users' inputs on touchscreen. In this paper, we reveal another motion accelerometer-based attack which is particularly serious: when a person takes the metro, a malicious application on her smartphone can easily use accelerator readings to trace her. We first propose a basic attack that can automatically extract metro-related data from a large amount of mixed accelerator readings, and then use an ensemble interval classier built from supervised learning to infer the riding intervals of the user. While this attack is very effective, the supervised learning part requires the attacker to collect labeled training data for each station interval, which is a significant amount of effort. To improve the efficiency of our attack, we further propose a semi-supervised learning approach, which only requires the attacker to collect labeled data for a very small number of station intervals with obvious characteristics. We conduct real experiments on a metro line in a major city. The results show that the inferring accuracy could reach 89\% and 92\% if the user takes the metro for 4 and 6 stations, respectively.
\end{abstract}


\section{Introduction}
Sensor-rich mobile devices such as smartphones and tablets have become ubiquitous. Ever-expanding users carry them everywhere. High-quality sensors (e.g., camera, GPS and accelerometer) on these devices continuously sense people-centric data and have helped developers create a wide range of novel applications. However, once these sensors are hijacked by malware, they may seriously threaten the user privacy. For instance, Templeman et al. \cite{TemplemanRCK13} recently introduce a visual malware that can exploit cameras on smartphones to construct rich, three dimensional models of users' homes or offices. Owusu et al. \cite{owusu2012accessory} find that accelerometers could be utilized to eavesdrop passwords that users input through touch screens.

While a good number of sensor-based threats have already been identified, this paper reveals a new one that is particularly serious. In brief, we find that if a person with a smartphone takes the metro, a malicious application on her smartphone can use the accelerometer readings to trace her, i.e., infer where she gets on and off the train. The cause is that metro trains run on tracks, making their motion patterns  distinguishable from cars or buses running on ordinary roads. Moreover, due to the fact that there are no two pairs of neighboring stations whose connecting tracks are exactly the same in the real world, the motion patterns of the train within different intervals are distinguishable as well. Thus, it is possible that the running of a train between two neighboring stations produces a distinctive fingerprint in the readings of 3-axis accelerometer of the mobile device, leveraging which attackers can infer the riding trace of a passenger.

We believe this finding is \textbf{especially threatening} for three reasons. First, current mobile platforms such as Andorid allow applications to access accelerometer without requiring any special privileges or explicit user consent, which means it is extremely easy for attackers to create stealthy malware to eavesdrop on the accelerometer. Second, metro is the preferred transportation mean for most people in major cities. For example, according to the Wikipedia, the daily of ridership of New York City Subway is between 2.5 million and 5.5 million, while that of Tokyo Metro is about 6.4 million. This means a malware based on this finding can affect a huge population. Last and the most importantly, metro-riding traces can be used to further infer a lot of other private information. For example, if an attacker can trace a smartphone user for a few days, he may be able to infer the user's daily schedule and living/working areas and thus seriously threaten her physical safety. Another interesting example is that if the attacker finds Alice and Bob often visit the same stations at similar non-working times, he may infer that Bob is dating Alice.

We emphasize that our attack is more effective and powerful than using GPS or cellular network to trace metro passengers. The first reason is that metro trains often run underground, where GPS is disabled. The second reason is that on most mobile platforms, applications have to request for user permissions before being able to access built-in localization components including both the GPS unit and the cellular localizer. In addition, while using these components, a particular icon usually appears on the screen, which will draw the attention of users soon. Therefore, we think that it is not a good choice to use built-in localization components to track metro passengers stealthily.

\medskip
\noindent\textbf{Methodology and Challenges}: There exist some nice dead reckoning mechanisms that exploit the smartphone accelerometer to estimate the moving directions and placements of a car \cite{NguyenZG10} or a walking man \cite{JinMSZ10, ZhuLC13}. So, the attacker may also leverage these mechanisms to reconstruct the train trajectory and then map it to the metro lines on the map to trace the passenger. Nevertheless, compared with walking and other transportation means, the running of metro trains is much gentler even at turning points, which means that their accelerometer readings are much smaller and more sensitive to even tiny noises. Consequently, the above methods designed for cars or humans, which require to precisely extract fine-grained micro information (e.g., turn angles, displacement) hidden in every few seconds of acclerometer readings, are not suitable for metro trains. According to our experiment, the predicted trajectory is far from the real one.

However, although fine-grain micro information is hard to learn, tens of seconds of accelerometer readings between between each pair of neighboring stations (called station interval) must expose some coarse-grained but easy to extract macro features (e.g., sharp peaks and valleys, amplitude variances at different directions) due to the track difference. Our methodology aims to extract such macro features from the accelerometer readings of every station interval and use machine learning techniques to learn interval classifiers, which are then used to detect stations that a specific passenger has passed. However, this task confronts the following challenges:

First, \emph{the metro readings are hidden in the data corresponding to other scenarios such as motionless, walking and taking other transportation means}. We need an appropriate method to extract metro readings accurately. Second, \emph{the metro features can be easily interfered by noises due to intentional or unintentional movements of users}. As a result, many station intervals may be falsely recognized. We need a robust trace inferring method that can tolerate recognition errors of individual station intervals. Third, \emph{it is too expensive for the attackers to collect sufficient labeled training data for every station interval in a large-scale metro system}. Namely, we cannot use supervised learning to learn interval classifiers.

\medskip
\noindent\textbf{Our Contributions}: we make the following specific contributions in this paper:

(1) We are the first to propose an accelerometer-based side channel attack for inferring metro-riders' traces. Our basic attack consists of two phases. In the training phase, the attacker collects labeled accelerometer readings for each station interval and extracts carefully-selected features to learn a set of interval classifiers. In the attack phase, malware installed on users' smartphones will automatically read and upload accelerometer readings. The attack first leverages the sharp amplitude difference between the data of metro and other transportation means to precisely extract metro-related data from miscellaneous accelerometer readings of a victim. It then segments this data by identifying brief stops and applies the interval classifiers to map data segments to station intervals. In this process, we use ensemble techniques to improve the classification accuracy of individual segments. Moreover, we leverage the fact that the translated intervals should be continuous to devise a voting-based trace inferring algorithm, which is able to further tolerate recognition errors of individual segments due to various noises.

(2) Since collecting labeled training data for each station interval in advance is impractical, we propose an improved attack that only requires the attacker to collect labeled data from a very small set of station intervals with obvious characteristics (e.g., the distance is much longer than the average, or with obvious turns). In particular, we devise a semi-supervised learning approach that is able to learn interval classifiers by combining this limited labeled data with a large amount of unlabeled data obtained from victims' phones in the attack phase.

(3) We conduct real experiments on Nanjing metro line 2 to evaluate the effectiveness of the proposed attack. We develop an Android application that can read the accelerometer data. Eight volunteers carry smartphones with this application installed when taking the Metro. Their traces cover 400 station intervals in total. The results show that the averaging inferring accuracy can reach about 70\% and 90\% when a volunteer rides the train for 4 stations and for 6 stations, respectively.

(4) In order to protect the location privacy of metro riders, we discuss several possible countermeasures against the attack we propose.

\section{Basic attack using supervised\\ learning}
\label{basic_attack}
This section presents a basic version of the proposed side-channel attack for tracing metro riders. This version requires the attacker to collect enough amount of labeled accelerometer data for each station interval (In this paper, a station interval refers to the track segment between two adjust stations) during the training phase, which is obviously impractical for a large-scale metro system. \textbf{We will describe how to avoid such supervised learning in the next section}.

\subsection{Attack Overview}
The proposed attack assumes that an attacker has infected a large number of users' smartphones with a carefully-designed malicious application. This application intermittently reads accelerometers and the orientation sensors, which are available on almost all the major mobile devices, and uploads the readings to remote servers through any available wireless networks. Such malware is not hard to create as both the accelerometers and the orientation sensors can be accessed without the authorization of users. Internet access needs the permission of users. Nevertheless, since almost every application applies for this permission, most users just grant without any hesitation. Besides the developing, the malware distribution is also easy to achieve based on existing social engineering mechanisms. Therefore, we will not focus on these two tasks in this paper.

The major goal of the proposed attack is to infer users' metro-ride traces, i.e., at which stations they get on and off, based on the metro-related data hidden in the collected sensor readings. The basic idea behind this attack is that the track differences among different \emph{station intervals} lead to different macro motion characteristics, which may be captured by the motion sensors (e.g., the accelerometers) of passengers' smartphones. As a result, it is possible for the attacker to extract these characteristics by analyzing the sensor readings and then utilize classic machine learning algorithms to identify the passengers' ride intervals.

\begin{figure}[!htb]
\centering
\includegraphics[width=3.5in]{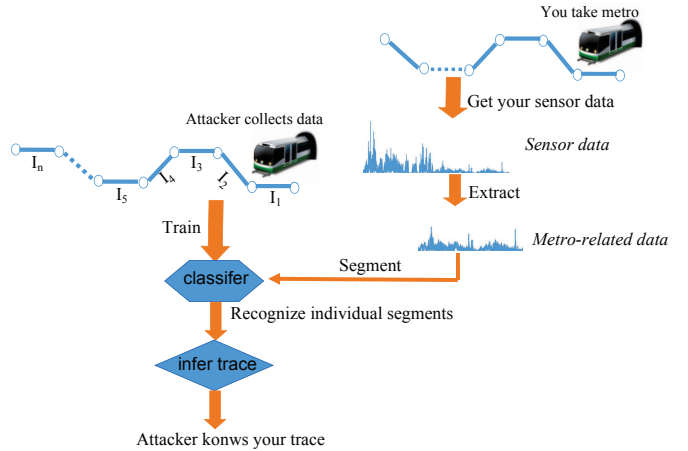}
\caption{attack model}
\label{attackmodel}
\end{figure}

As we show in Fig.~\ref{attackmodel}, the proposed attack is composed of two phases. In the training phases, the attacker collects motion sensor readings for each station interval and then uses a supervised learning scheme to build an interval classifier. In the recognition phase, the attacker analyzes the sensor readings collected by the malware from infected smartphones and then utilizes the interval classifier to identify the station intervals that users pass by. Specifically, this phase contains the follow three key steps:

(1) \textbf{Metro-related data extraction}: Among the large amount of data collected by the malware, only a small proportion of it is corresponding to the metro riding. Most of it is generated when the users stay still, walk or take other means of transportation. On account of this, we need first solve the challenge to filter out metro-related data from the mixed sensor readings.

(2) \textbf{Data segmentation and recognition}: As station intervals are the basic recognition primitives for the classifier constructed in the training phase, we need further segment the metro-related data for each user. Each data segment is corresponding to one station interval. We achieve this goal by searching for the stop slots of trains, in which accelerometer readings are smaller than other areas. Then, the attacker applies the interval classifier to map these data segments to station intervals.

(3) \textbf{Metro-ride trace inferring}: Although the previous step maps the segments data to the specific station intervals, the recognition results might be contradictory with each other because of errors. For example, two neighboring segments of data are mapped to non-neighboring intervals. Targeting this problem, we present a voting based algorithm to infer the complete metro-ride trace of a user by taking all his segment recognition results into consideration.

\subsection{Coordinate Transformation}
\begin{figure}[!htb]
\centering
\includegraphics[width=3.0in]{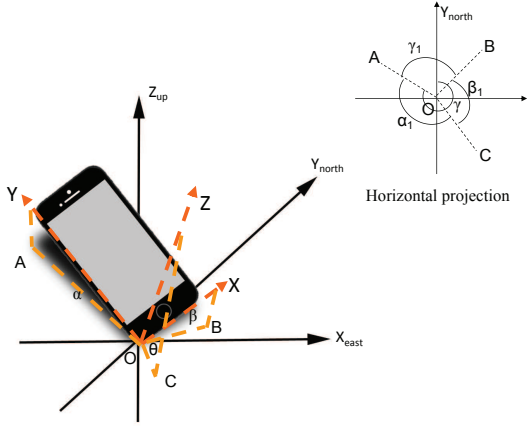}
\caption{Coordinate decomposition of the sensor}
\label{coordinate}
\end{figure}

The proposed attack uses the readings of 3-axis accelerometers on smartphones to infer metro passengers' traces. As we show on Fig. \ref{coordinate}, each reading is a three-dimensional vector $[x,y,z]$ in a screen-based dynamic coordinate system ($\vec{X}$,$\vec{Y}$,$\vec{Z}$), which rotates as the phone rotates. So this system varies from phone to phone. Thereby, it is hard to derive any meaningful motion patterns of metro trains from raw readings. To solve this problem, we introduce another static East-Noth-Up (ENU) coordinate system which is also shown in Fig. \ref{coordinate}. This system does not rotate as the phone rotates. We thereby transform every reading $[x,y,z]$ in the original phone system to $[x',y',z']$ in ENU system before performing any analysis.

It is impossible to directly perform this transformation due to the lack of the relation between these systems. We should use the orientation sensor, which is a virtual sensor based on the magnetometer, to achieve this goal. The reading of the orientation sensor is also three-dimensional data $[\alpha,\beta,\gamma]$, where $\alpha$ is the angle between the $Y$-axis with respect to the horizontal plane, and $\beta$ is the angle of the $X$-axis and the horizontal. $\gamma$ is the angle between the horizontal projection of the $Y$-axis of the phone system and true north, With these three angles, it is easy to derive the east, the north and the up components of the acceleration (i.e. the vector $[x',y', z']$ in ENU coordinate system). We show the results in Table~\ref{tab1}. The angles $\gamma_1$, $\alpha_1$, $\beta_1$, $\theta$ are marked in Fig.~\ref{coordinate}.

\begin{table}[!htb]
\centering
\caption{Results of coordinate transformation}
\scalebox{0.8}{
\label{tab1}{
    \begin{tabular}{|p{1.0cm}|p{3.0cm}|p{4.5cm}|}
        \hline

         ECA & the east component of the acceleration  & $y\cos\alpha\cos(\gamma-\pi)+x\cos\beta\cos(\gamma+\gamma_1-\pi)+z\cos\theta\cos(\gamma+\beta_1-\pi)$ \\
        \hline
         NCA & the north component of the acceleration & $-y\cos\alpha\cos(\gamma-\pi/2)-x\cos\beta\cos(\gamma+\gamma_1-\pi/2)-z\cos\theta\cos(\gamma+\beta_1-\pi/2)$ \\
        \hline
         VCA & the vertical component of the acceleration & $x_1\sin\beta+y_1\sin\alpha+z_1\sin\theta$ \\
        \hline
    \end{tabular}
}
}
\end{table}

\subsection{Extraction of metro related data}
\label{emrd}
After the coordinate transformation, the next task for the attacker is to extract metro-related data from a large amount of sensor readings collected from victim smartphones. Among these readings, only a small fraction are produced when users take the metro. Hemminki et al. \cite{hemminki2013accelerometer} propose an elegant accelerometer-based transportation mode detection mechanism on smartphone. We may directly apply this proposal to fulfil our task. However, their goal is to achieve fine-grained detection of the transportation means for each piece of accelerometer data, which is much more complex than ours, i.e., to precisely determine whether a give piece of accelerometer data corresponds to metro or not. Thus, we devise a simpler solution for this challenge.

\begin{figure}[!htb]
\centering
\includegraphics[width=3.0in]{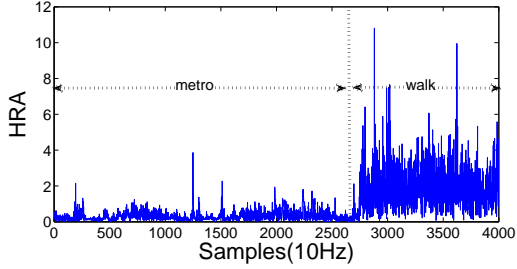}
\caption{The horizontal accelerations of walking and taking the metro}
\label{walkandmetro}
\end{figure}

\begin{figure*}[!htb]
        \subfigure[Bus]{
                \includegraphics[width=3.0in]{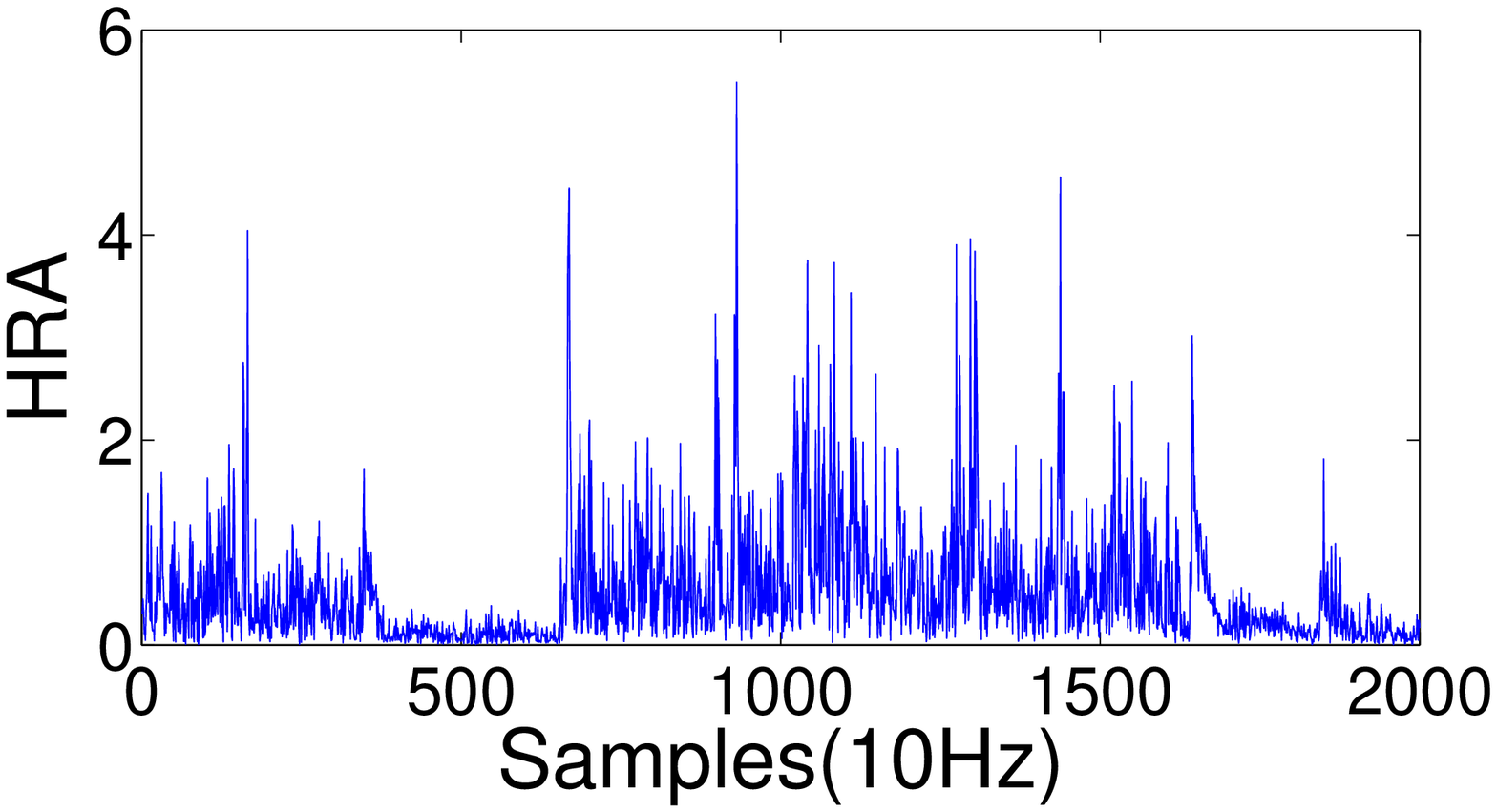}
                \label{fig:bus}
        }%
        \subfigure[Taxi]{
                \includegraphics[width=3.0in]{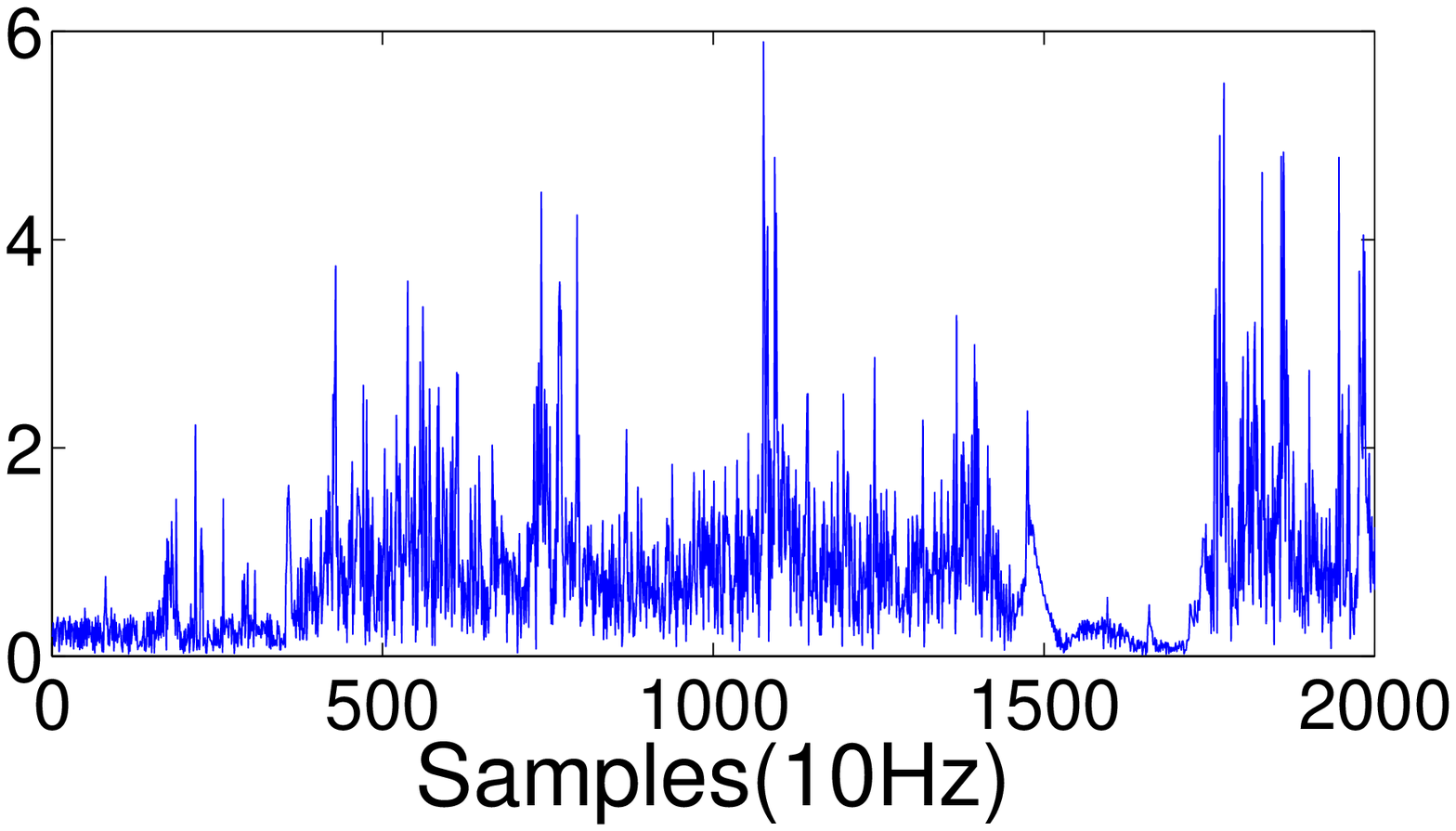}
                \label{fig:taxi}
        }
        \subfigure[Metro]{
                \includegraphics[width=3.0in]{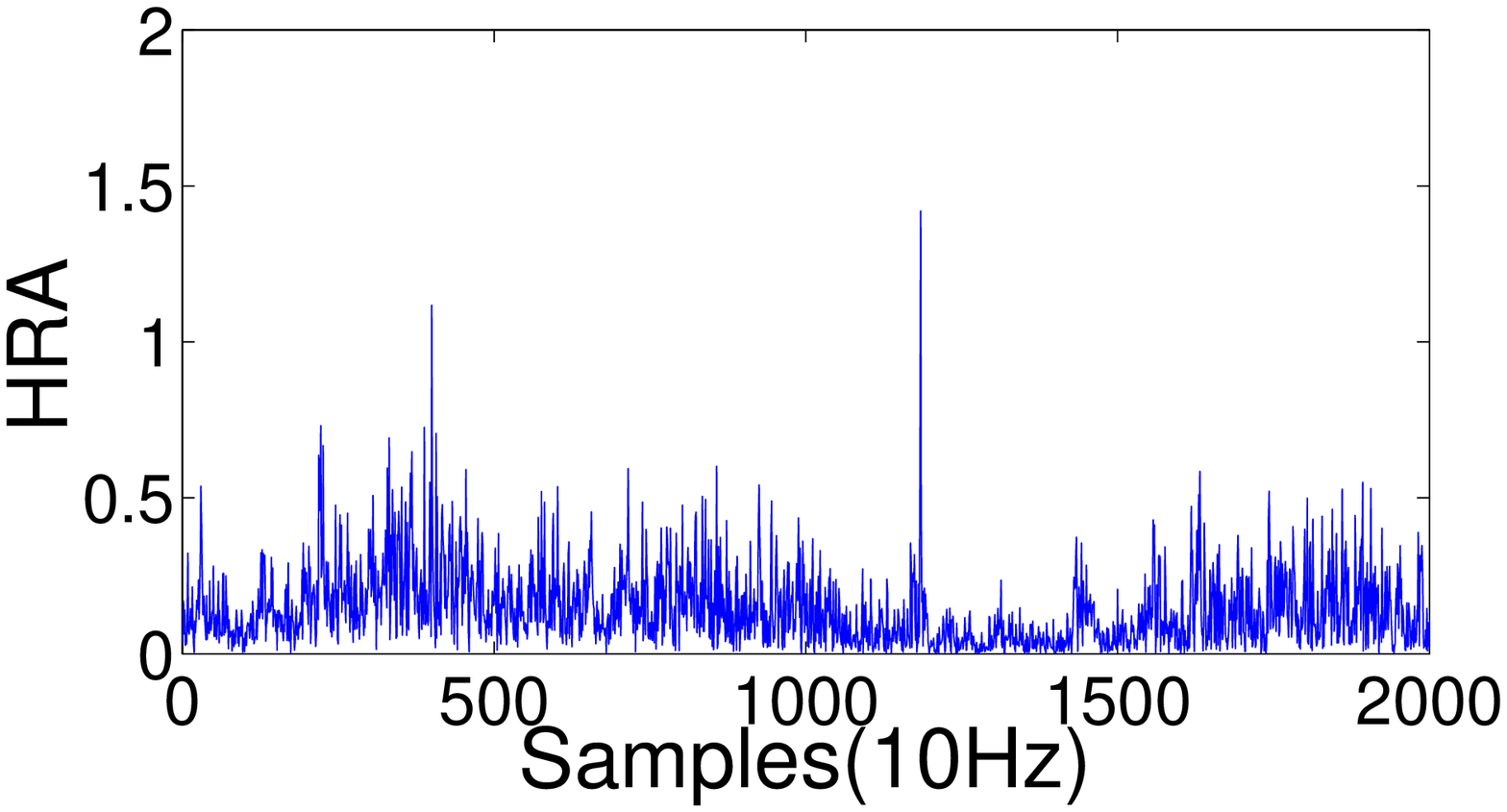}
                \label{fig:metro}
        }
         \subfigure[Static]{
                \includegraphics[width=3.0in]{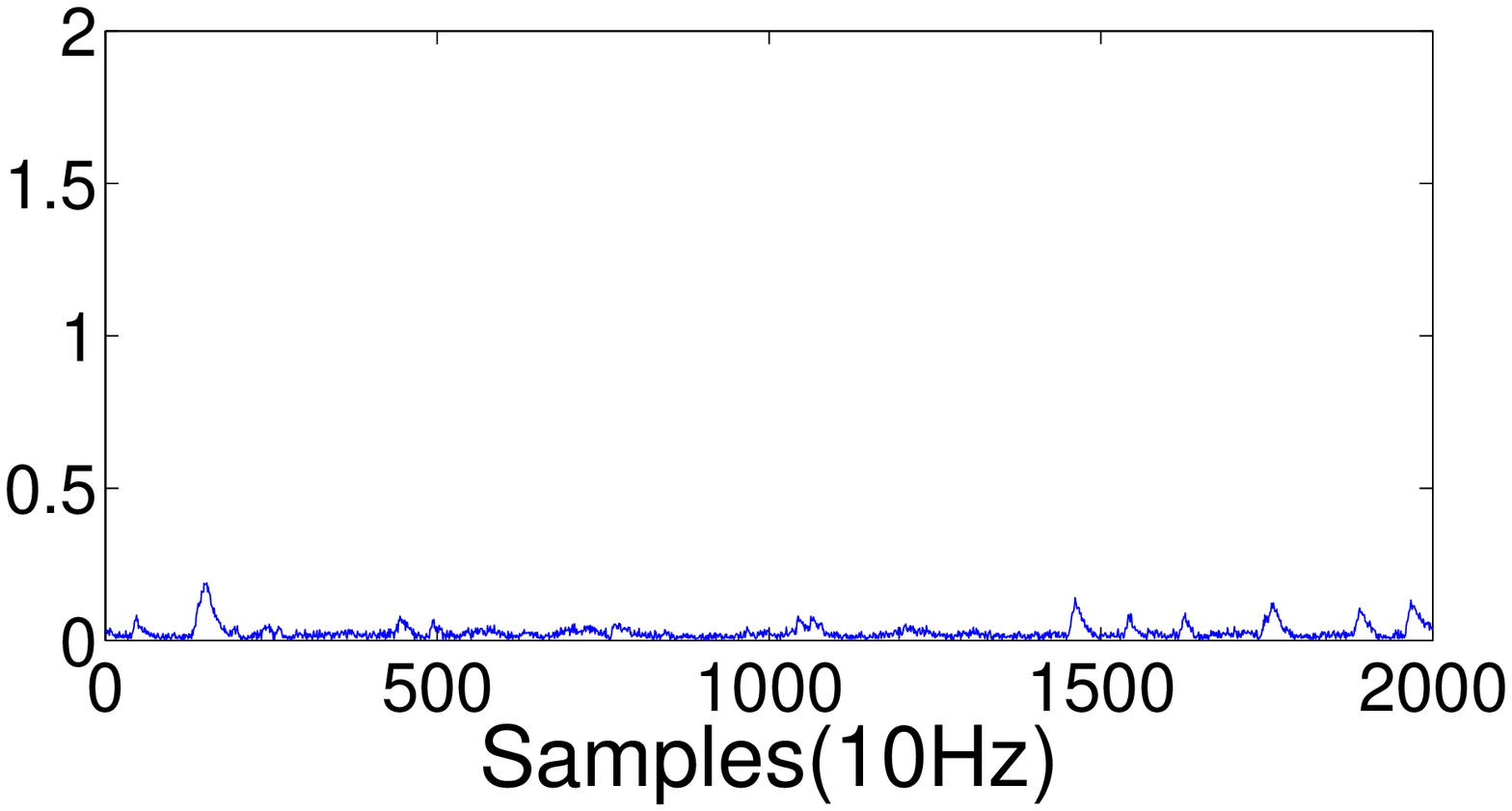}
                \label{fig:static}
        }
        \caption{The horizontal acceleration of traveling by bus, taxi, metro and static, respectively}
        \label{fig:transportation}
\end{figure*}

To extract metro-related data, we have to first learn the distinction between it and the data related to other transportation means. Fig.~\ref{walkandmetro} presents the sequential values of the horizontal resultant acceleration (HRA) when a user changes from metro to walk. Left is the data generated on the metro, while right is the data corresponding to walk. We can find that the amplitude of the walk-related data is significantly larger than that of the metro-related. Fig.~\ref{fig:transportation} further compares the HRA curves when the user is on the metro, taxi and bus. We can still observe a sharp difference that the amplitude of metro data is much smaller than that of the non-metro data.

Based on the above observations, we build a naive bayes classifier based on the HRA charactersitics to identify metro-related data from mixed sensor readings. Given a sequence of HRA values of a victim, the attacker classifies each $m$-sample sliding window. We use five statistical measures of the HRA values: mean, variance and the numbers of samples that surpass three pre-defined thresholds, respectively, as the classification features. The classification result is binary: either \emph{metro} or \emph{non-metro}. We move the window $m$ samples in each sliding. The window size is set to be the half of the length of the shortest station interval in the target metro network. The last feature is picked because we think it can well capture the amplitude difference between metro and other transportation forms in our observation.

According to our experiments in \refsec{exp:ext}, this simple classifier may produce errors, especially the false positives. However, we observe that it is rare to find two consecutive widows that are both misclassified. This is because the length of the classification window is usually longer than one minute, which is not very short. It is unlikely for other transportation means to move the same as a metro train more than two minutes. We thereby propose the following optimization to further reduce the errors.

If $Win_i$ is classified as \emph{non-metro} while $Win_{i+1}$ is classified as \emph{metro}, we first continue to classify $Win_{i+2}$. If it is attributed to \emph{non-metro}, we think that $Win_{i+1}$ is misclassified. Otherwise, a new sequence of metro data is considered to begin at some position within $Win_i$. In this case, we further classify the windows beginning at Sample $(i+1)w-1$,$(i+1)w-2$,$\cdots$ one by one until meeting the first window that is classified to be \emph{non-metro}. Then, if the start position of this window is at Sample $(i+1)w-k$ ($1<k\leq w$), the start position of the new sequence of metro data is considered to be at Sample $(i+1)w-k+w/2$. We can use a similar method to handle the suitation if $Win_i$ is \emph{metro} while $Win_{i+1}$ is \emph{non-metro}. Due to the space limitation, we omit the detailed description here.

\subsection{Segmentation of metro related data}
\label{std}
After obtaining the metro-related data, the next step is to segment it and let each segment correspond to a station interval. We segment the data because station intervals are the recognition primitives for the interval classifier built in the training phase.
\begin{figure}[!htb]
\centering
\includegraphics[width=3.0in]{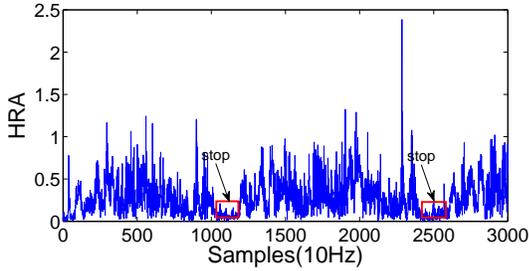}
\caption{Illustration of stop slots}
\label{split}
\end{figure}

As we know, metro train has to make a brief stop between any two station intervals for disembarking and loading passengers. We thereby try to segment the metro-related data of a victim by identifying the stop slots hidden in the data. Fig. \ref{split} shows the HRAs derived from a piece of metro-related data. We can observe that there exist a series of periodic slots where the values are much smaller than those of other positions. According to our analysis, these slots rightly correspond to the stop periods of the train. The values in these slots are smaller because the train is still and has no acceleration in any direction. We design an algorithm shown in \ref{findfinalsegmentpoint} to automatically determine the segmenting points by searching for this kind of stop slots.

Let $\{X_1,X_2,\cdots,X_n\}$ be a sequence of HRA values derived from a victim's metro-related data. The proposed algorithm defines a sliding window $W$, the length $L_{W}$ of which is equal to the minimum time of a brief metro stop. It moves forward $W$ from $X_1$ one value by one value until reaching a sample $X_i$ that the number of values below a threshold $T_1$ within $W_{X_i}$ exceeds $95\% L_{W}$, i.e.,
$$
|\big\{k\in\{i,i+1,i+L_{W}-1\}:X_k<T_1\big\}|>95\% L_{W}.
$$
Then, we regard $X_s$ ($s\in{i,i+1,\cdots,i+win/2-1}$) that minimizes $Mean(W_{X_s})$ as a potential segmenting point. Here, $Mean(W_{X_s})$ is defined to be the mean value of the points within $W_{X_s}$. Once we find $X_s$, the algorithm directly skips the next $T_2$ values and searches for the next stop slots from $X_{s+T_2}$. Here, $T_2$ equals the length of the shortest station interval in the target metro system.

\begin{algorithm}
\caption{FindFinalSegmentPoints}
\label{findfinalsegmentpoint}
\SetKwFunction{FindSegPoints}{FindSegPoints}\SetKwFunction{SizeOf}{SizeOf}
\SetKwInOut{Input}{Inputs}\SetKwInOut{Output}{Output}
\Input{A sequence of HRA values of a victim's metro-related data, $X$;\\A threshold for identifying stop slots, $T_1$;\\
The maximum length of a station interval, $L_{max}$;\\
The minimum length of a station interval, $L_{min}$\\}
\Output{Final segmenting points;}
\BlankLine
\Begin{
$orderedSet$ =  \\\qquad\FindSegPoints{$(X,0,Length(X), T_1)$}\;
$isStop$ = $false$\;
\While{!$isStop$}
{
    $isStop=false$\;
    OrderSet $tmpSet=\emptyset$\;
    $T_1=T_1+\Delta$\;
    \For{$i\leftarrow 0$ \KwTo \SizeOf{$set$}}
    {
        \If{$set[i+1]-set[i]>L_{max}$}
        {
            $isStop=false$\;
            $tmpSet$ $\bigcup=$\\\quad\FindSegPoints{$(X, set[i]+$\\\qquad $L_{min}, set[i+1]-L_{min},T_1)$}\;
        }
    }
    $set=set\bigcup tmpSet$\;
}
\KwRet{$set$}
}
\end{algorithm}
The above process could help us identify a set of potential segmenting points. However, sometimes due to selecting an unsuitable $T_1$, it may miss one or several segmenting points, especially when the sensor data contains many noises (Note that, false segmenting points can be avoided by making $T_1$ small enough ). To address this problem, we further check the segmenting points that we just find. If the distance between neighboring points goes beyond the maximum length of a station interval, we know that some segmenting points between them must have been missed. So, the algorithm slightly increases $T_1$ and re-searches the stop slots within that interval. To improve the accuracy, we repeat this step until the distance between any two adjacent segmenting points does not exceed the maximum interval distance.

According to our experiments, this approach may still produce some errors even after applying the above measure. So in Section. \ref{ti} we will give a further solution to tolerate erros in the trace inferring.
\begin{procedure}
\caption{FindSegPoints()}
\SetKwInOut{Input}{Inputs}
\SetKwInOut{Output}{Output}
\Input{$X$, $T_1$, $L_{max}$, $L_{max}$; \\The start and end index of $X$: $sIdx$ and $eIdx$;}
\Output{A set of potential segmenting points;}
\Begin
{
    $i=0$\;
    OrderedSet $retSet=\emptyset$;
    \While{$i<eIdx-L_{min}$}
    {
        $count=|\{k\in W_{X_i}:X_k<T_1\}|$;
        \If{$count>80\%L_{W}$}
        {
            Find $s\in\{i,i+1,\cdots, i+win/2-1\}$ that minimizes $Mean(W_{X_s})$\;
            $retSet$ = $\bigcup= \{X_s\}$\;
            $i += L_{min}$\;
        }
        \Else{
            $i++$\;
        }
    }
    \KwRet{$retSet$}
}
\end{procedure}
\subsection{Recognition of the stations}
\label{rts}
By now we have discussed how to segment the metro-related data. In this section we will further discuss how to distinguish among data segments.
Our basic attack requires the attacker to collect sufficient amount of training data for each station interval (We will introduce how to bypass this limitation in \refsec{iasl}). It then utilizes the labeled data to lean a classifier model, which helps translate the data segments returned in the last step to the station intervals. We now first detail the feature selection and then introduce the classification approach.

\subsubsection{Feature Selection}
The features used for classification can be divided into two sets.

(1) \textbf{Statistical Features}
As we show in \reftab{tab2}, this set includes statistical features of the accelerometer data of the target segment in both time and frequency domains. Note that we extract these features for all 3 individual components in \reftab{tab1}, which indicates that the total number of features in this set reaches 24. These features are able to effectively capture overall patterns of the train movement during this interval. For instance, the STD of the NCA component is useful to characterize the vertical vibration pattern of the train. Note that before extracting these features, we first perform the signal smooth that we will describe soon to filter out random noises due to the movements of the user hands.
\begin{table}[!htb]
\center
\caption{Statistical features that we used for classification}
\label{tab2}
\scalebox{0.8}
{
\begin{threeparttable}
\centering
    \begin{tabular}{|c|c|}
        \hline
        Mean & Means of acceleration\\
        \hline
        Max & Maximum of acceleration\\
        \hline
        STD & Standard deviation of acceleration\\
        \hline
        MAV & Mean of the absolute value of acceleration\\
        \hline
        NVHT1, 2, 3 & Number of values higher than threshold1, 2, 3\\
        \hline
        Length & Length of the segment\\
        \hline
        FFT DC 1-6 Hz & Six first FFT components\\
        \hline
        SE & Spectral Entropy\\
        \hline
        SP & Spectrum peak position\\
        \hline

    \end{tabular}
\end{threeparttable}
}
\end{table}

(2) \textbf{Peak Features}
Although statistical features can capture overall patterns of the train movements, they may miss some local significant events such as big turns at particular positions, which are usually caused by significant changes of the metro track and are ideal features for the interval classification. These events usually result in the sharp peaks and valleys in the accelerometer data. So, to capture such critical features, we include the top three peaks and valleys of accelerations on each axis in NEU system in the feature vector. Nevertheless, these features are not easy to extract.

\begin{figure}[!htb]
\centering
        \subfigure[Original noisy data]{
                \includegraphics[width=2.0in]{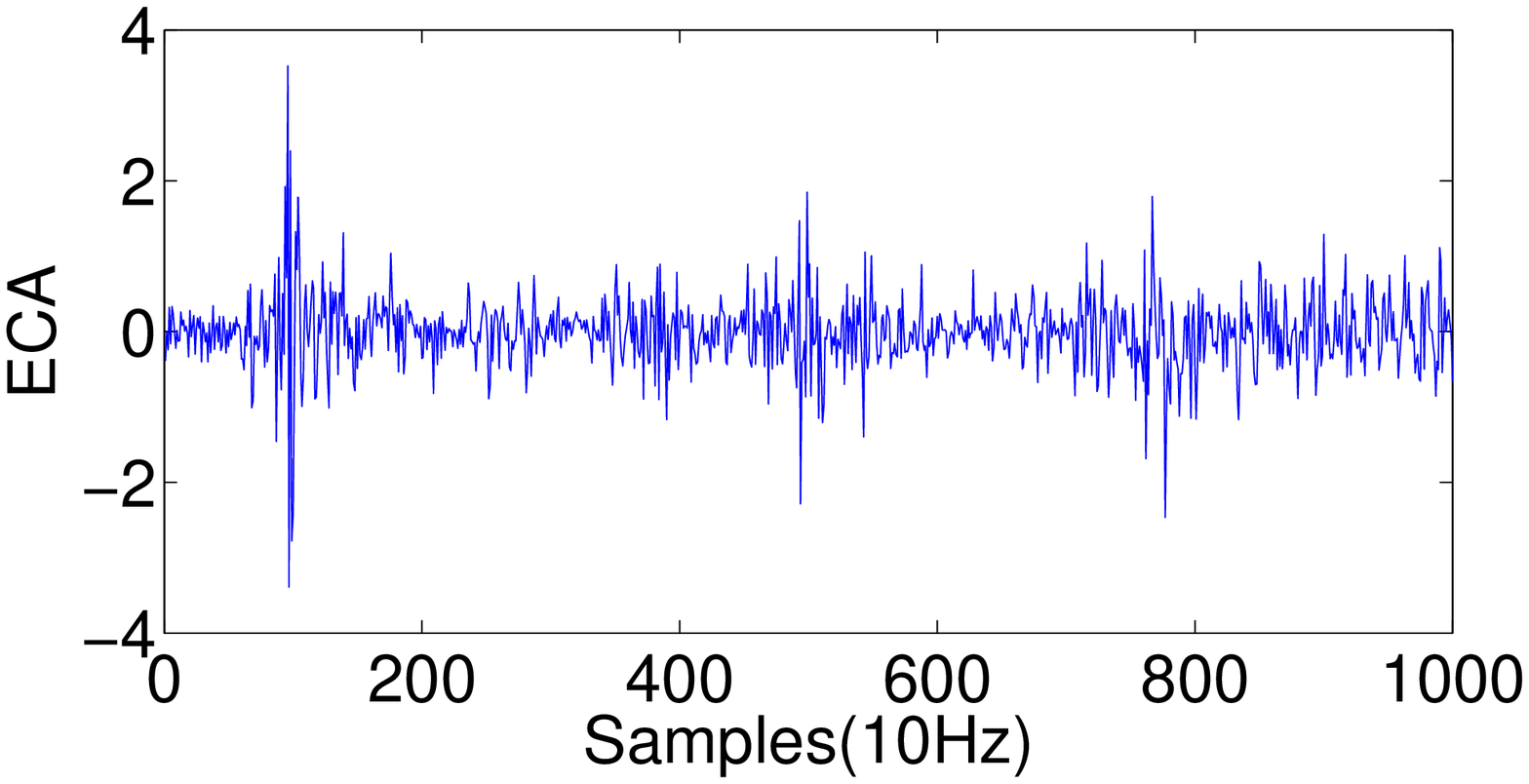}
                \label{fig:waggle}
        }
        \subfigure[Smoothed data]{
                \includegraphics[width=2.0in]{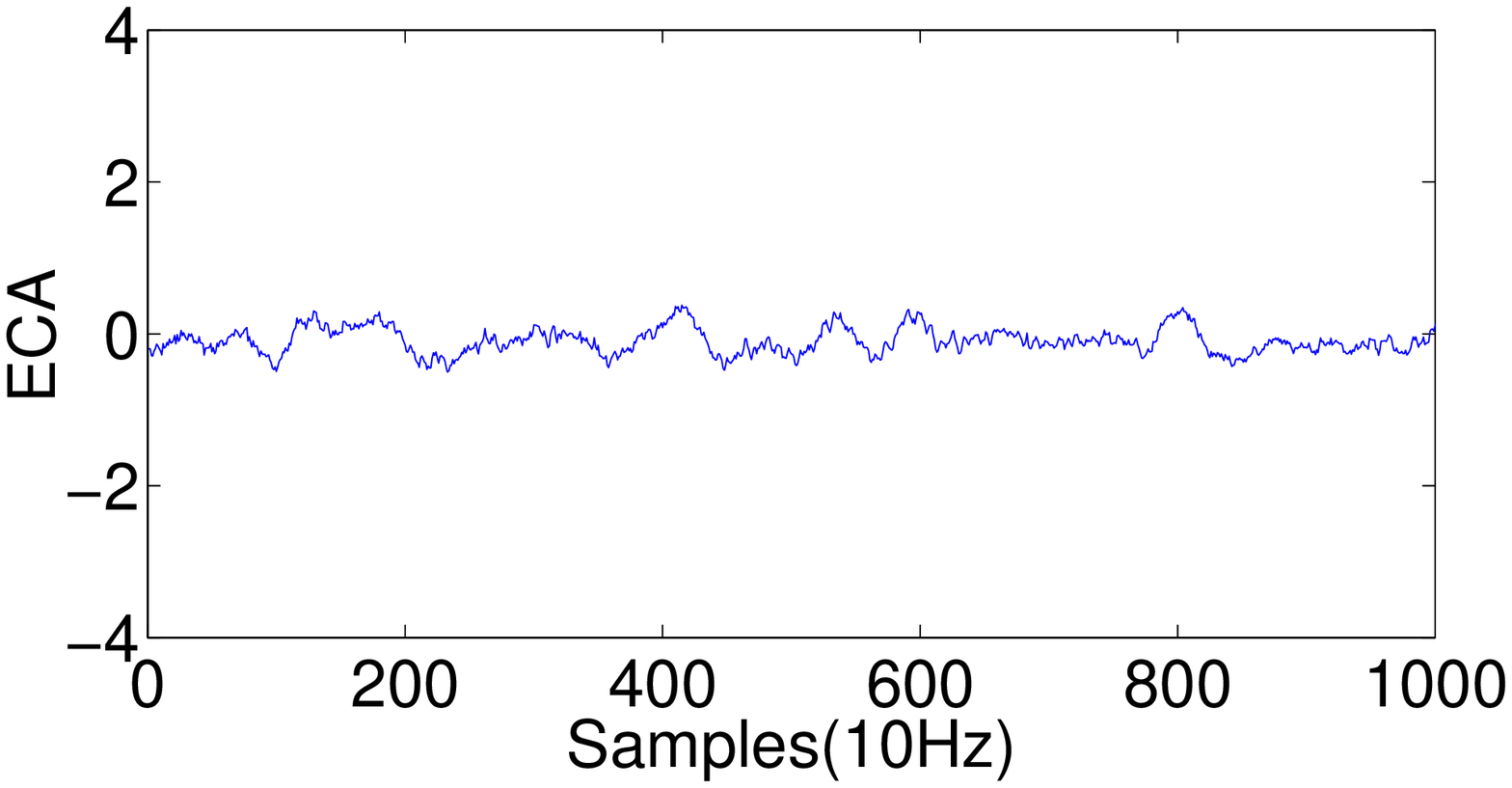}
                \label{fig:waggles}
        }
        \caption{The effect of smooth}
        \label{fig:smooth}
\end{figure}

First of all, the accelerometers data may include many noises due to the hand movements of the user. Fig.~\ref{fig:waggle} shows the accelerations on the east axis when a user shakes his hand holding the smartphone in the stationary case. We can find that this shaking may produce larger peak or valley amplitudes than the movement of metro trains. We employ a simple smooth technique to reduce the interference of such noises. Specifically, for an acceleration sample $X_i$ on a specific axis, this technique replaces its value with the average of the samples within a $k$-sample window around it, i.e.,
$$X_i'=Average(X_{i-2/k},X_{i-2/k+1},\cdots, X_{i+2/k-1}).$$
We present the accelerations after being smoothed in Fig.~\ref{fig:waggles}. We can find that the amplitudes of the new curve become much more smaller, and their peaks and valleys can hardly interfere the extraction of desired features now. This technique works since the accelerations due to the hand movement will change from one direction to the opposite in a short-term, and thus the sum can cancel each other. The accelerations due to the train movement, however, may last for a longer time in one direction, and will not cancel each other.

\begin{figure}[!htb]
\centering
\includegraphics[width=3.0in]{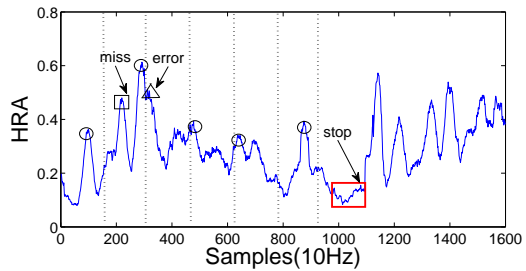}
\caption{Illustration on peak selecting }
\label{waveridge}
\end{figure}

Second, according to our experiments, we find that one significant change of the metro track may cause multiple random peaks or valleys that are extremely close. As they can only reflect a single feature of the station interval, it is better to avoid including all of them into the feature vector for the classification. Thus, as shown in Fig.\ref{waveridge}, we divide a specific acceleration curve into windows of the same size, find and rank the maximum (minimum) value in each window, and regard the three top ranking maximums (minimums) as the desired peaks (valleys). Nevertheless, if the window size is set improperly, this approach may still make mistakes. For instance, the windows in Fig.\ref{waveridge} not only miss a desired peak, but also find a false peak. To further improve the accuracy, we repeat the above process serval times with different window sizes, and chooses the three peaks (valleys) that win the most times as the final outputs.

\subsubsection{Classification}
After we determine the features, we use them extracted from the labeled data to train a classifier for recognizing unknown data segments. Instead of using only one classification model, we train multiple basic multi-class classifiers and use the ensemble technique \cite{dietterich2000ensemble,breiman2001random,freund1995desicion} to combine the classification results with the aim of creating an improved composite classifier. The final class prediction is based on the votes of the basic classifiers. We mainly use two types of classifiers: boosted Naive Bayesian and decision trees.

To improve the accuracy of Naive Bayesian, we implement its boosted version based on the AdaBoost \cite{freund1995desicion} algorithm. In AdaBoost, weights are assigned to each training tuple. A series of $k$ classifiers are iteratively learned. In each round of leaning, the samples from the original training set is re-sampled to form a new training set. The samples with higher weights are selected with a higher chance. After a new classifier $M_i$ is learned, the samples that are misclassified by $M_i$ are assigned higher weights, which makes the following classifier $M_{i+1}$ pay more attention to the misclassified tuples. The final prediction result is returned based on the weighted votes of the classifiers learned in each round. For the decision tree technique, we also use its ensemble version, random forests \cite{breiman2001random}, to improve the accuracy. In particular, this technique generates a collection of diverse decision trees by randomly selecting a subset of the features and training tuples for learning. During classification, each tree votes and the final result equally considers all these votes.

Although we have applied the ensemble technique to improve the classification, it is impossible to completely remove errors due to various kinds of noises existing in the training data. However, the trace of a passenger usually contains more than one segment, which should be translated to continuous station intervals. If some of them are misclassified, the translated results are very likely to become discontinuous, i.e., cannot form a practical passenger trace. This enables us to filter out some classification errors. On the other hand, this property also
indicates that we have the chance to obtain a correct trace so long as one of the elemental segments is correctly recognized. In the next subsection, we leverage this observation to propose a voting-based trace inferring mechanism which can better tolerate classification errors of individual segments.
\subsection{Error-Tolerant Trace Inferring}
\label{ti}
Assume the metro-related data of a passenger consists of $n$ segments $\{S_1,S_2,\cdots,S_n\}$, and the metro network contains $m$ station intervals $\{I_1,I_2\cdots,I_n\}$. Instead of returning the single winner, we make the classification mechanism proposed above return a probability matrix $\mathbf{P}=[P_{i,j}]_{n\times m}$ where $P_{i,j}$ denotes the probability that data segment $S_i$ is mapped to station interval $I_j$.

As the inferring result must be a continuous sequence of station intervals of length $n$, the inferring domain $\wp$ is actually limited. For instance, if we assume that the $m$ station intervals belong to one metro line, there are only $2\times(m-n+1)$ possible results: $\wp=\{I_1\rightleftharpoons I_n, I_2\rightleftharpoons I_{n+1},\cdots, I_{m-n+1}\rightleftharpoons I_{m}\}$. We can exhaustively consider each of these possibilities, and use a voting-based approach to determine the final output. In this approach, the votes that one possibility $Pbt_i$: $I_i\rightarrow\cdots\rightarrow I_{i+n-1}$ obtains equal the sum of the probabilities for each data segment to be mapped to the corresponding station interval in $Pbt_i$, i.e., $Vote(Pbt_i)=\sum\limits_{j=1}^nP_{j,i+j-1}$. We simply pick the possibility obtaining the highest votes as the final inferring result. This method can \emph{well tolerate classifying errors of individual segments} for two reasons:

(1) For each data segment, we take into account not only its optimal mapping but also other possibilities. Note that the optimal mapping may be incorrect due to classifying errors.

(2) Our final inferring result comprehensively considers the classification results of all the  member segments in a trace. The errors of one or a small number individual segments may not affect the overall predication.

When the metro system is large, we should reduce the size of $\wp$ to improve the efficiency of the above process. For this purpose, we pick three station intervals with the highest mapping probabilities for every data segment $S_i$. Then, we only include the possibility $I_{k-i+1}\rightarrow\cdots\rightarrow I_k\rightarrow \cdots I_{k+n-i}$ into $\wp$ for every such interval $I_k$. By doing so, the size of $\wp$ will be greatly reduced.

The accuracy of such kind of inferring heavily relies on the correctness of data segmenting. If the later is incorrect, the inferred outcome must be either wrong. Although we have taken some measures to increase the segmenting precision in \refsec{std}, some errors may still exist as we show in \reffig{accseg}. To tolerate such errors, if the user data is segmented into $n$ segments by the algorithm, we also consider the conditions of being segmented into $n-1$ and $n+1$ segments. Specifically, for every possibility $I_{k_1}\rightarrow\cdots\rightarrow I_{k_n}$ in the optimized $\wp$, we consider $I_{k_1}\rightarrow\cdots\rightarrow I_{k_{n-1}}$ and $I_{k_1}\rightarrow\cdots\rightarrow I_{k_{n+1}}$ as well. Note that the time length of a station interval can be estimated based on the map. So, when we compute the probability of $I_{k_1}\rightarrow\cdots\rightarrow I_{k_{n-1}}$, we can segment the user data into $n-1$ segments based on the estimated length of each interval.

\section{Improved attack using \\semi-supervised learning}
\label{iasl}
The basic attack proposed in the last section requires the attacker to collect labeled data for each station interval for building an interval classifier. However, in the real world, there are many cities, such as New York and Tokyo, which consist of tens of metro lines and hundreds of station intervals. It is extremely time consuming for the attacker to traverse every station by metro many times. In this section, we aim to address this problem by proposing an improved attack using semi-supervised learning to significantly reduce the workload of the attacker.
\begin{figure}[!htb]
\centering
\includegraphics[width=2.5in]{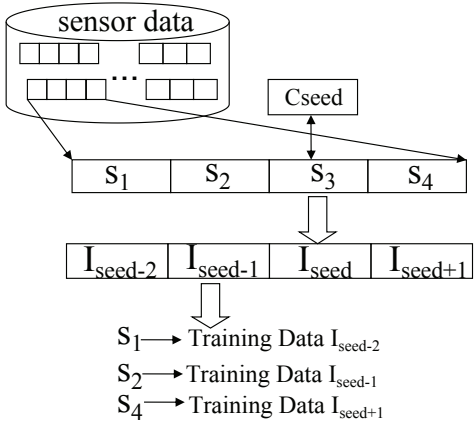}
\caption{One round of semi-supervised learning}
\label{orsl}
\end{figure}

\begin{algorithm}
\caption{Proposed Semi-supervised learning for labeling user data}
\label{speculateSection}
\SetKwInOut{Input}{Inputs}\SetKwInOut{Output}{Output}
\SetKwFunction{Identify}{Identify} \SetKwFunction{Training}{Training}
\Input{Lists of unlabeled data segments, $SLists$;\\
The seed classifier,$C_{seed}$;
}
\Output{Lists of labeled segments, $Result$}
\BlankLine
\Begin{
    $CSet = \{C_{seed}\}$\;
    \While{True}{
        $LLists = \emptyset$\;
        \ForEach{$SL\in SLists$}
        {
            $IL$ = \Identify{$SL$, $CSet$}\;
            \For{$i=0$; $i<IL.length$; i++}
            {
                $LLists[IL[i]]\leftarrow SL[i]$\;
            }
        }
        $count=0$\;
        \ForEach{$LL\in LLists$}
        {
            \If{$LL.length>Threhold$}
            {
                $count++$\;
                $C=$ \Training{LL}\;
                $CSet[C.ID]= C$\;
            }
        }
        \If{$count ==$ Total Number of Intervals}{break\;}
    }
    \Return $LLists$\;
}
\end{algorithm}

In the improved attack, the attacker is only required to personally collect sensor data for one or a very small number of station intervals with obvious features, e.g. containing big turns, which can guarantee a high recognition rate. It tries to use these intervals as the seeds to infer unlabeled data belonging to other intervals.

Without loss of generality, we assume that the attacker only collects labeled data for a single station interval, which is denoted by $I_{seed}$. The overview of the proposed semi-supervised learning algorithm is present in \refalg{speculateSection}. It first builds a particular binary classifier $C_{seed}$ for $I_{seed}$ based on the corresponding labeled data. This classifier uses the same set of features in \refsec{rts} and returns a binary result that whether an input segment corresponds to $I_{seed}$ or not. Similar as the classification method in \refsec{rts}, the classifier here may be an ensemble combines a series of basic classifiers. Next, it uses this classifier to check the segmented unlabeled data collected from victims. If a segment of a victim's data sequence is classified as $I_{seed}$ \, we can easily infer the belongings of other segments in the same sequence. For instance, if $S_3$ in the sample sequence $<S_1S_2S_3S_4>$ is classified as $I_{seed}$, we know that $S_1$, $S_2$ and $S_4$ are mapped to $I_{seed-2}$, $I_{seed-1}$ and $I_{seed+1}$, respectively. It then labels these data segments and adds them to the training sets of the corresponding station intervals. After finishing checking a large number of victims' data, we may have obtained enough number of labeled training data for some non-seed station intervals. So, we can build particular binary classifiers for these intervals as well. In the next round, we treat these intervals as new seeds, and use them to classify the victims' data segments again. In this round, some intervals that do not get enough training data in the last round may get enough data now, and therefore can be regarded as new seeds.

We repeat the above process until all the station intervals get enough training data. By now, we can turn back to the basic attack. The difference is that all the training data except that of the seed interval are produced by inferring instead of being personally collected by the attacker. This may reduce the accuracy of the final trace inferring, but not significantly according to our experiments.

Note that due to classifying errors, different seed classifiers may produce contradictory results. For instance, consider a victim's data sequence $<S_1S_2S_3S_4>$. Suppose that in a specific round the attacker has obtained two seed classifiers $C_i$ and $C_j$. If $C_i$ recognizes $S_i$ as $I_i$, $C_j$ recognizes $S_2$ as $I_j$, but $I_i$ and $I_j$ are not continuous, we get an conflict. This problem can be solved based on a similar voting-based method as that in \refsec{ti}. Specifically, the classification result of each seed classifier is regarded as a vote. We finally return the result receives the highest votes. Here, each vote is weighted according to the classification confidence.
\section{Experiment}
\label{Experiment}
In this section, we introduce our experiments on real metro for evaluating the feasibility of the proposed attack.

\begin{figure}[!htb]
\centering
\includegraphics[width=3.5in]{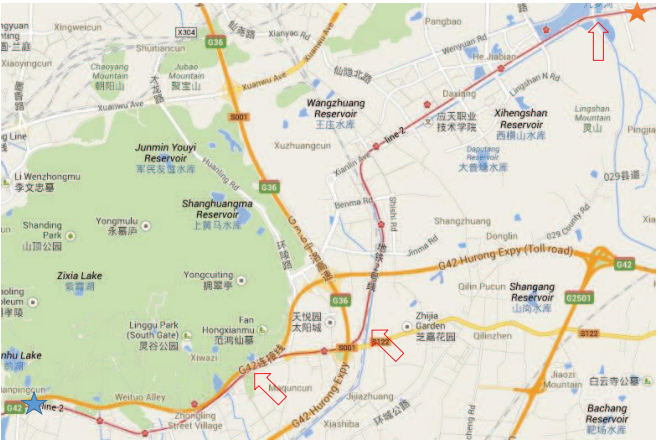}
\caption{Map of the metro line used in our experiment}
\label{map}
\end{figure}

Our experiments are performed on Nanjing metro line 2. Fig.~\ref{map} shows the range of the metro line. Eight volunteers repeatedly travel between two stations by metro. Each of them carries an Android smartphone that installs a data-gathering application developed by us. This application reads the accelerometer and the orientation sensor every 0.1s, and automatically uploads the accumulated data to a remote server when WiFi is available. During experiments, smartphones are held in hands, and the testers are operating them in usual ways. The phones that testers used include Samsung S3, S4 and Note2. We finally collect forty data sequences, each of which corresponds to a trip containing 10 station intervals. So the dataset covers $400$ data segments in total. We then evaluate the effectiveness of the proposed attack based on this dataset.

\subsection{Accuracy of the extraction of metro-related data }\label{exp:ext}
\begin{figure}[!htb]
\centering
        \subfigure[Before optimization]{
                \includegraphics[width=2.5in]{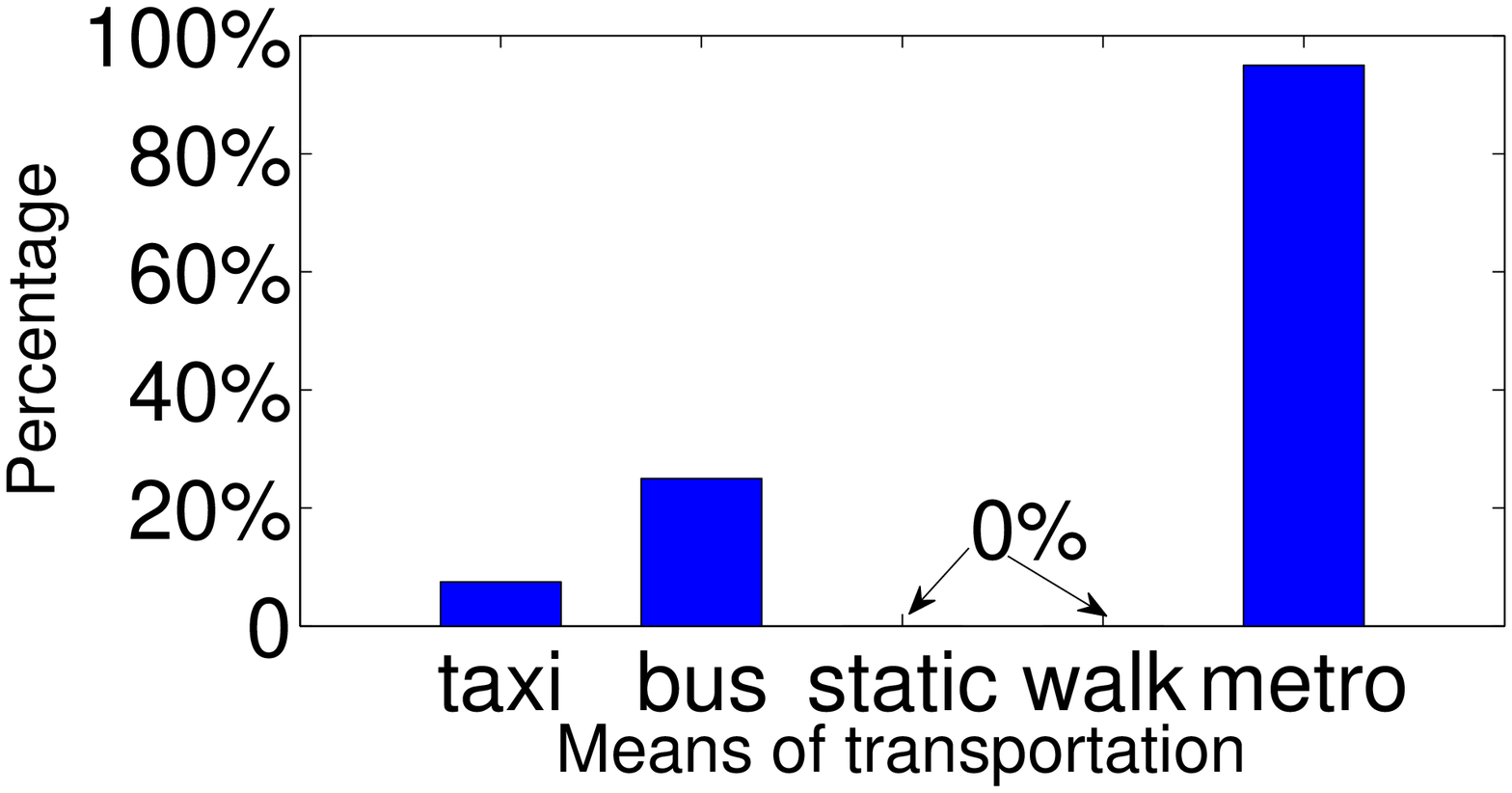}
                \label{recognizebefore}
        }
        \subfigure[After optimization]{
                \includegraphics[width=2.5in]{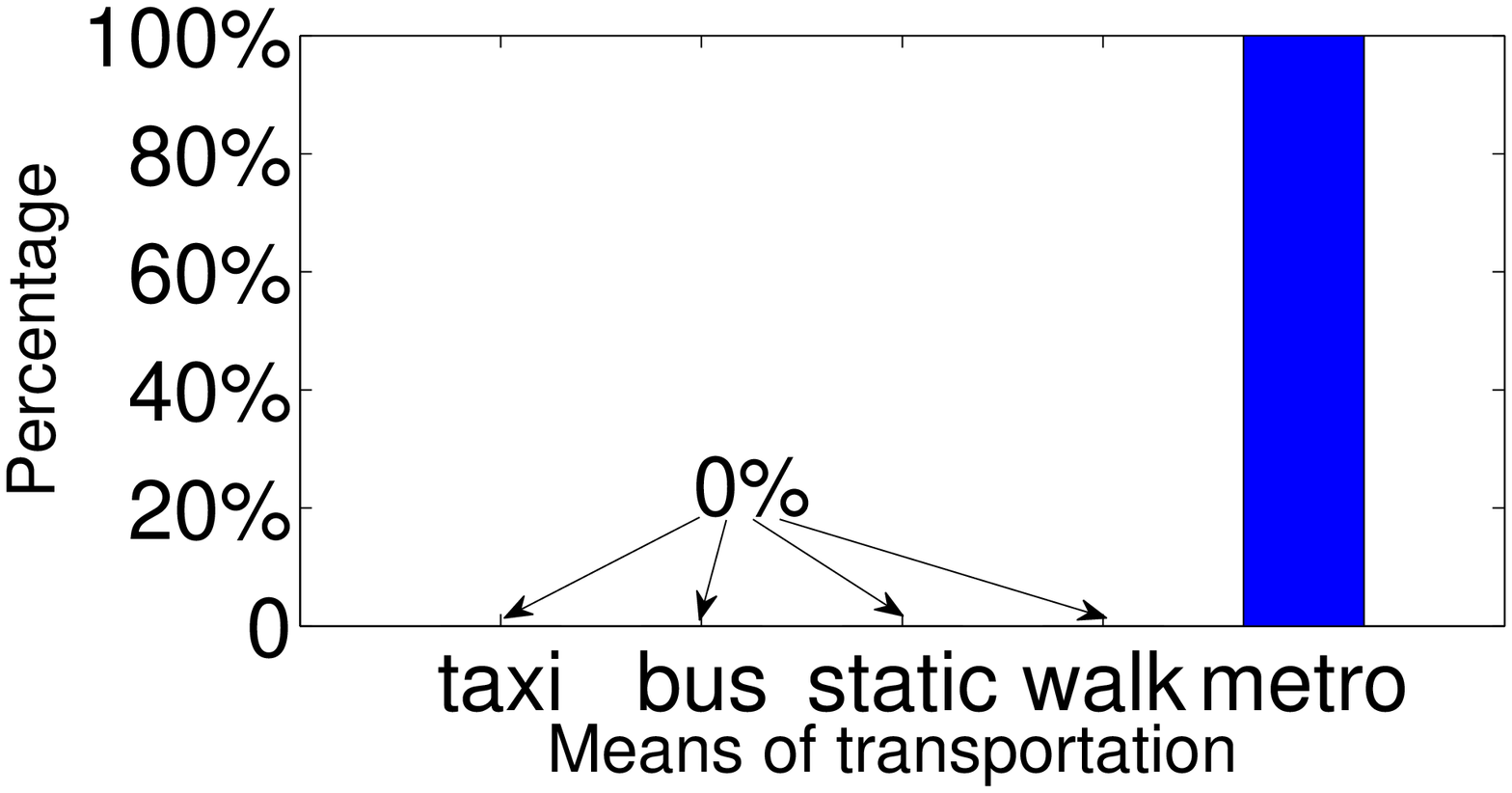}
                \label{recognizeafter}
        }
        \caption{Accuracy of extracting metro-related data}
        \label{fig:recognize}
\end{figure}

We first evaluate the accuracy of our method for extracting metro-related data. For this purpose, besides the metro-related data, we also collect 1.5h data each for four other transportation means include walking, bus, taxi and stillness. We divide each of these data sequences (including 40 metro-related data sequences) into 100 second-long segments, and then use the classifier that we introduced in Sec.~\ref{emrd} to classify them. The percentage for each kind of data to be classified as metro-related is presented on Fig.~\ref{recognizebefore}. We can find there will be some errors in the extraction. But when we use the further optimization that we proposed in the end of Sec.~\ref{emrd}, more than 99 $\%$ of the metro data is correctly recognized, and no false positives are produced.

\subsection{Accuracy of segmenting the metro-related data}\label{accseg}
\begin{figure}[!htb]
\centering
\includegraphics[width=3in]{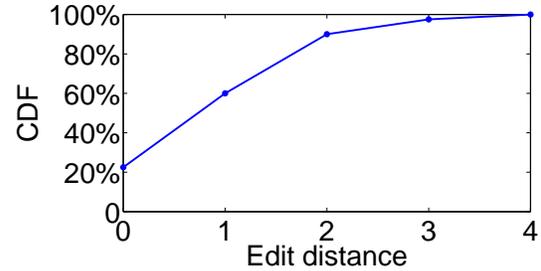}
\caption{Segmenting accuracy measured by the edit distances between the segmenting results and the facts}
\label{editdistance}
\end{figure}

We now evaluate the accuracy of our method in \refsec{rts} to segment the metro-related data. We employ \emph{Edit Distance}, which is a popular way of quantifying the dissimilarity between two strings, to measure the segmenting accuracy. Suppose that $A=X_{j_1}X_{j_2}\cdots X_{j_n}$ is the real sequence of segmenting points of a victim's metro data, while the counterpart produced by Algorithm 1 is $B=X_{k_1}X_{k_2}\cdots X_{k_m}$. The edit distance $ED(A,B)$ is defined to be the minimum number of operations required to transform $B$ into $A$. Here, different from in the string scenario, we assume that two nodes, $X_{j_s}$ and $X_{k_t}$, are equal so long as $|j_s-k_t|<10s$, where 10s is half of the minimum stop-time of the trains. We segment every data sequence in our experimental dataset, and the CDF of the edit-distance distribution is presented in \reffig{editdistance}. We can find that more than 90$\%$ segment sequence which compared to the real sequence the error point is less than 2. We thereby should employ the mechanism propose in the end of \refsec{ti} to tolerate these errors in the trace inferring.
\subsection{Accuracy of the basic attack}
\label{taba}
In this subsection, we evaluate the inferring accuracy of the basic attack using supervised learning. As we mentioned earlier, we totally collect forty groups of metro-related data, each of which corresponds to a 10-station-interval trip. In each evaluation, we pick $39$ of $40$ sequences for training, leaving one for testing. We do not vary the ratio of training and testing data here because this attack is just the basic version. In our subsequent evaluation on the improved attack, all these 40 data sequences are regarded as unlabeled testing data.

We first evaluate the classification accuracy of the naive Bayes classifier. The results are shown in Table.~\ref{tab4}. The cell at row $i$, column $j$ denotes the percentage for the data segments corresponding to station interval $I_i$ to be classified as $I_j$.  We can find that the values at the diagonal positions are the greatest in most of the rows, which is a desired feature because these values equal the accurate recognition rates of station intervals. In addition, Table.~\ref{tab4} shows that station intervals $I_1$, $I_6$ and $I_7$ posses higher recognition rates than others. By checking these intervals on the map, we find that this result is reasonable because all these three intervals expose remarkable characteristics in their tracks that can significantly improve recognition rates.
\begin{table}[!htb]
\centering
\caption{Mapping probability from individual segments to station intervals($\%$)}
\label{tab4}
\scalebox{0.8}{
   \begin{tabular}{|c|c|c|c|c|c|c|c|c|c|c|}
        \hline
                 &$I_1$&$I_2$ &$I_3$ &$I_4$ & $I_5$ & $I_6$ & $I_7$& $I_8$&  $I_9$ & $I_{10}$ \\
         \hline
           $S_1$ &\multicolumn{1}{>{\columncolor{mygray}}l}{80}& 0 & 5 & 0 & 2.5 & 5 & 5&  0&  0 & 2.5\\
        \hline
           $S_2$ & 2.5 & \multicolumn{1}{>{\columncolor{mygray}}l}{40} & 5&  5&  10 & 0 & 0 & 22.5&  12.5 & 2.5\\
        \hline
           $S_3$ & 2.5&  10  &\multicolumn{1}{>{\columncolor{mygray}}l}{47.5} & 7.5 & 7.5 & 5 & 10 & 0 & 7.5 & 2.5\\
        \hline
          $S_4$ & 15 & 10&  7.5  &\multicolumn{1}{>{\columncolor{mygray}}l}{45} & 0  &0 & 0  &10 & 8 & 5\\
        \hline
         $S_5$ & 5& 12.5 & 2.5 & 10 &\multicolumn{1}{>{\columncolor{mygray}}l}{37.5} & 2.5 & 17.5 &5 & 2.5&  5 \\
        \hline
          $S_6$ & 0 & 0 & 5 & 0 & 5 & \multicolumn{1}{>{\columncolor{mygray}}l}{80} & 10 & 0 & 0 & 0\\
        \hline
          $S_7$ & 5 & 2.5 & 0 & 0 & 2.5 & 12.5 & \multicolumn{1}{>{\columncolor{mygray}}l}{77.5} & 0&  0 & 0\\
        \hline
          $S_8$ & 7.5 & 32.5 & 10 & 2.5 & 7.5&  0 & 2.5 & \multicolumn{1}{>{\columncolor{mygray}}l}{22.5} & 12.5 & 2.5\\
        \hline
          $S_9$ & 5 & 15 & 2.5 & 7.5 & 7.5 & 2.5 & 0 & 15 & \multicolumn{1}{>{\columncolor{mygray}}l}{37.5} & 7.5\\
        \hline
         $S_{10}$ & 10 & 22.5 & 7.5 & 5 & 10 & 5 & 10 & 7.5 & 2.5 &  \multicolumn{1}{>{\columncolor{mygray}}l}{20}\\
        \hline
    \end{tabular}
    }
\end{table}

We then evaluate the performance of our voting-based inferring mechanism proposed in \refsec{ti}. In each round of evaluation, we still pick 39 of forty sequences for training. For the remained one containing 10 segments, we slip it and generate three sets of subsequences, whose lengths are 3 (segments), 5 and 7, respectively. Each sequence in these sets is considered as the metro-related data of a distinct passenger. We then apply the voting-based method to infer his trip. The inferring accuracies for the sequences of different lengths are presented in \reffig{infer}. We find that the inferring accuracy increases with the length of the data sequence, i.e., the trip length of the passenger. Specifically, when the trip is composed of 3 station intervals, the average inferring accuracy is about 80\%. When the length is increased to 7, this value is greater than 94\%.

\subsection{The accuracy of the improved attack}
Finally, we evaluate the inferring accuracy of the improved attack using semi-supervised learning. In our experiment, we pick $I_6$ and $I_7$ as the seed intervals since their recognition rates are relatively higher (Please see Table.~\ref{tab4}) thanks to their obvious characteristics. We mark them by read arrows in fig.~\ref{map}. We collect additional 20 pieces of training data for each of them and then construct classifiers for them separately using supervised learning. In this case, the 40 data sequences are all unlabeled and regarded as testing data. In addition, because, in the real world, unlabeled data collected from passengers usually has varied lengthes, we randomly split these data sequences. Each subsequence of a random length is considered as one piece of independent training data. We then run the mechanism proposed in Sec.~\ref{iasl} to infer training data for each station interval. Once all the station intervals get sufficient number of training data, we can return to perform the basic attack against the testing data.

\begin{figure}[!htb]
\centering
\includegraphics[width=3.0in]{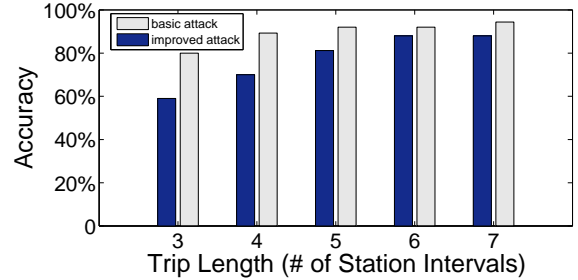}
\caption{Final inferring accuracy on the traces}
\label{infer}
\end{figure}

The average inferring accuracies for the data of different lengthes are presented in Fig.~\ref{infer}. We can find that compared with those in Fig.~\ref{infer}, all the results decrease, but not so significantly. The inferring accuracies for the trips of length 3, 5 and 7 can surpass 59\%, 81\% and 88\%, respectively.

\subsection{Power Consumption}
\begin{figure}[!htb]
\centering
\includegraphics[width=3.5in]{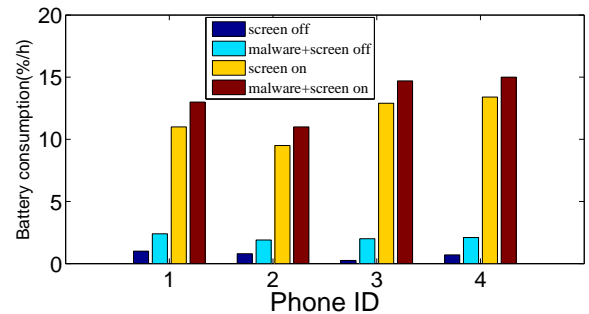}
\caption{Power Consumption}
\label{power}
\end{figure}

Malware in our attack has to continually access the accelerometer, which certainly consumes additional power. We thereby have performed a coarse-grained evaluation of power consumption of the application we have used in the above experiments. Note that we do not consider the power consumed for uploading the recorded data through WiFi since this operation is performed infrequently. As we show in Fig.~\ref{power}, we compare the power consumptions when our background application is running or not on four different smartphones. We consider both the scenarios that the screen is on and off. Phone ID 1 to 4 correspond to Samsung S4, Huawei G750, Samsung S3, MEIZU MX, respectively. We can find that the increased power consumption per hour (less than 1.8\%) due to the running of this application is quite limited .

In our current malware implementation, it keeps reading the accelerometer every 0.1s once being launched. We can optimize this design by moving the task of extracting metro data from the server to the smarphone. If it detects that the current accelerometer data does not correspond to metro, it can sleep for a longer time (e.g., 5min). It is of small probability for the user to take metro in the short time if he is not taking metro now. By doing so, the above energy consumption is expected to be further reduced.
\section{Scheme of defense}
Our scheme does not rely on GPS or other positioning systems, which gives it a high level of concealment and considerable efficiency. It may disguise itself into normal smartphone software when it steals the information on the users' trace. However, some defensive strategies can be made to lower the chance of the leakage of information on users' trace.

(1) The smartphones that we currently use does not inform the user that the application will need the permission to access to the sensors. To prevent the leakage of the users¡¯ privacy, we let the operating system hint the user that the application will access to the sensors and ask for users' permission. However, this step is usually neglected by most users.

(2) We may blend some noise into the sensor data in order to prevent attackers from making use of sensor data to grab the users' privacy effectively. If the user needs the original sensor data without noise, selection dialog boxes will prompt out to let users permit the use of non-noise sensor data to some applications. This may ensure that the privacy of users will not leak by sensor data.

(3) If malware intends to steal the users' privacy through sensor data, constant request for the data from sensors will evidently boost the power consumption. No matter how the malware tries to conceal itself, the acquisition of sensor data will lead to an increasing power comsumption of the smartphone. We may scrutinize the status of power consumption of programs to examine those programs that keep consuming too much electricity. In this way, it is highly possible for us to find malware that operates background.
\section{Related Work}
In our work, we dig information from metro-related sensor data. Actually there are many works in which users¡¯ private data are stolen through accelerometer. Liu et al. \cite{liu2009uwave} design a software called uWave. It makes use of the triaxial accelerometer in a smartphone to recognize the gestures of the users, which has achieved a good effect. Wu et al. \cite{wu2009gesture} also do some research on gesture recognition. So if malicious attackers utilize those data, they will know what users do. Cai et al. \cite{cai2011touchlogger} initial a project in which he reckons the users¡¯ taps on their smartphones by accelerometers. As taps on different places of the smartphone screen will bring different changes to the sensor, given the fixed arrangement of smartphone keyboard, password and other personal information that users have typed out may be revealed. Compared with brute-force attack, this approach is much more effective \cite{owusu2012accessory}.

Our work infers the metro-related sensor data, there are also works that involve the data from accelerometer as a part of the whole database in order to trace the user. Lee and Mase \cite{lee2002activity} point out that motor data can be used to speculate on the users' traces, but a starting point needs to be settled first. However, accelerometer is just one of the sensors that are utilized and this work is not based on smartphone. Han et al. \cite{han2012accomplice} use accelerometers of smartphone only to deduce users' trace. They acquire the users' tracks by the algorithm they design, then they match them with a map to infer the trace of the user, which has inspired us a lot. We implement their model which called ProbIN. We experiment on the metro line, but it can not draw the metro line accurately. It can hardly recognize the turns of the metro line, we can only draw a straight line without turns. Their experiment is based on driving, As we can see from Fig.~\ref{fig:transportation}, the metro-related data is smooth, so it is sensitive to noise, it can hardly get detail information about metro from this method.

Our work involves extracting metro-related data from a lot of sensor data, which inevitably takes the recognition of different means of transportation into consideration. The earliest ways to recognize the form of transportation is based on a multi-sensor platform \cite{bao2004activity,consolvo2008activity}. As smartphones develop, some early systems use embedded accelerometers to read out traveling on foot or other non-motorized means of transportation, such as walking \cite{miluzzo2008sensing,iso2006gait}, running, ascending and descending the stairs \cite{brezmes2009activity} or riding the bicycles \cite{bieber2009activity}. Some of the works have achieved a good effect on recognizing those means, whose accuracy is higher than 90\%. There are also some works about detecting stationary and motorised transportation modalities \cite{reddy2010using,wang2010accelerometer}. But the result yielded is much less effective \cite{hemminki2013accelerometer} compared to detecting of non-motorized transportation modalities. Hemminki et al. \cite{hemminki2013accelerometer} raise an advanced method, which largely improve the accuracy of the recognition of electrified transportation. They determine the user takes what kind of transportation. But in our work, we only need to determine if the user takes metro or non-metro transportation. So it is unnecessary for us to use this powerful but sophisticated method to extract metro-related data. In our work we propose a simple but equally effective algorithm to achieve our goal.

\section{Conclusion}
In this paper, we have proposed a basic attack which can extract metro-related data from mixed acceleration readings, then use an interval classier built from supervised learning to infer users' trace. This attack need the attacker to collect labeled training data for each station interval, so we further proposed a semi-supervised learning approach. The improved attack only needs to collect labeled data for a few station intervals with obvious characteristics.

We conduct real experiment on Nanjing metro line 2. From the experiment in Sec. \ref{Experiment} we find that the inferring accuracy can reach $92\%$ if the user takes the metro for 6 stations.

\bibliographystyle{IEEEtran}
\bibliography{IEEEabrv,bibtex}

\end{document}